# Discovery of Web Usage Profiles Using Various Clustering Techniques

Zahid Ansari[1], Waseem Ahmed[2], M.F. Azeem[3] and A.Vinaya Babu[4]
[1,2]Dept. of Computer Science Engineeing
P.A. College of Engineering
Mangalore, India
[1]zahid.ansari@acm.org

[3]Dept of Electronics and Communication
P.A. College of Engineering
Mangalore, India

[4]Dept. of Computer Science Engineering
J.N.T. University
Hyderabad, India

*Abstract*— The explosive growth of World Wide Web (WWW) has necessitated the development of Web personalization systems in order to understand the user preferences to dynamically serve customized content to individual users. To reveal information about user preferences from Web usage data, Web Usage Mining (WUM) techniques are extensively being applied to the Web log data. Clustering techniques are widely used in WUM to capture similar interests and trends among users accessing a Web site. Clustering aims to divide a data set into groups or clusters where inter-cluster similarities are minimized while the intra cluster similarities are maximized. This paper reviews four of the popularly used clustering techniques: *k*-Means, *k*-Medoids, Leader and DBSCAN. These techniques are implemented and tested against the Web user navigational data. Performance and validity results of each technique are presented and compared. (Abstract)

*Keywords-component; web usage mining; k-means clustering; k-medoids clustering; leader clustering; DBSCAN*

I. INTRODUCTION

Web Usage Mining [1] is described as the automatic discovery and analysis of patterns in web logs and associated data collected as a result of user interactions with Web resources on one or more Web sites. The goal of Web usage mining is to capture, model, and analyse the behavioural patterns and profiles of users interacting with a Web site. The discovered patterns are usually represented as collections of URLs that are frequently accessed by groups of users with common interests. Web usage mining has been used in a variety of applications such as i) Web Personalization systems [2], ii) Adaptive Web Sites [3][4], iii) Business Intelligence [5], iv) System Improvement to understand the web traffic behaviour which can be utilized to decide strategies for web caching [6], load balancing and data distribution [7], iv) Fraud detection: detection of unusual accesses to the secured data [8], etc.

Clustering techniques are widely used in WUM to capture similar interests and trends among users accessing a Web site. Clustering aims to divide a data set into groups or clusters where inter-cluster similarities are minimized while the intra cluster similarities are maximized. Details of various clustering techniques can be found in survey articles [9]-[11]. The ultimate goal of clustering is to assign data points to a finite system of k clusters. Union of these clusters is equal to a full dataset with the possible exception of outliers. Clustering groups the data objects based only on the information found in the data which describes the data objects and the relationships between them.

Some of the main categories of the clustering methods are [12]: i) *Partitioning* methods, that create k partitions of a given data set, each representing a cluster. Typical partitioning methods include k-means, k-medoids etc. In *k-means* algorithm each cluster is represented by the mean value of the data points in the cluster called centroid of the cluster. On the other hand and in *k-medoids* algorithm, each cluster is represented by one of the data point located near the center of the cluster called medoid of the cluster. Leader clustering is also a partitioning based clustering techniques which generates the clusters based on an initially specified dissimilarity measure, ii) *Hierarchical* methods create a hierarchical decomposition of the given set of data objects. A hierarchical method can be classified as being either agglomerative or divisive, based on how the hierarchical decomposition is formed. iii) *Density- based* methods form the clusters based on the notion of density. They can discover the clusters of arbitrary shapes. These methods continue growing the given cluster as long as the number of objects or data points in the "neighborhood" exceeds some threshold. They can also filter out noise and outliers. DBSCAN is a typical density-based method that grows clusters according to a density-based connectivity analysis. iv) *Grid-based* methods quantize the data object space into a finite number of cells that form a grid structure. All the clustering operations are performed on the grid structure. v) *Model-based* methods, that discover the best fit between data points given a mathematical model. Mathematical model is usually specified as a probability distribution.

The remainder of the paper is organized as follows. Section II presents a overview of web usage mining using clustering techniques and the underlying concepts. Section III presents each of the *k*-Means, *k*-Medoids, Leader and DBSCAN clustering techniques in detail along with the underlying mathematical foundations. Section IV describes the experimental results of each technique, followed by a comparison of the results. A brief conclusion and future work are presented in Section V.

## II. WEB USAGE MINING USING CLUSTERING

A number of clustering algorithms have been used in Web usage mining where the data items are user sessions consisting of sequence of page URLs accessed and interest scores on each URL page based on the characteristics of user behaviour such as time elapsed on a page or the bytes downloaded [2]. In this context, clustering can be used in two ways, either to cluster users or to cluster items. In user-based clustering, users are grouped together based on the similarity of their web page navigational patterns. In item based clustering, items are clustered based on the similarity of the interest scores for these items across all users. Mobasher et. al. [13], [14] have used both user-based clustering as well as item-based clustering in a personalization framework based on Web usage mining.

A typical user-based clustering starts with the matrix representing the user sessions or user profiles and partitions this multi-dimensional space into k groups of profiles that are close to each other based on a measure of distance or similarity among the vectors (such as Euclidean or Manhattan distance). Clusters obtained in this way can represent user segments based on their common navigational behaviour or interest shown in various URL items. In order to determine similarity between a target user and a user segment represented by the user session clusters, the centroid vector corresponding to each cluster is computed which is the representation of that user segment. To make a recommendation for a target user u and target URL item i, a neighbourhood of user segments that have a interest scores for i and whose aggregate profile is most similar to u are selected. This neighbourhood represents the set of user segments of which the target user is most likely to be a member. Given that the aggregate profile of a user segment that contains the average interest scores for each item within the segment, a prediction can be made for item i using k-nearest-neighbor approach [15].

We map the user sessions as vectors of URL references in a *n*-dimensional space. Let $U = \{u_1, u_2, ..., u_n\}$ be a set of *n* unique URLs appearing in the preprocessed log and let $S = \{s_1, s_2, ..., s_m\}$ be a set of *m* user sessions discovered by preprocessing the web log data, where each user session $s_i \in S$ can be represented as $s = \{w_{u_1}, w_{u_2}, ..., w_{u_m}\}$. Each $w_{u_i}$ may be either a binary or non-binary value depending on whether it represents presence and absence of the URL in the session or some other feature of the URL. If $w_{u_i}$ represents presence of absence of the URL in the session, then each user session is represented as a bit vector where

$$w_{u_i} = \begin{cases} 1; & \text{if } u_i \in s; \\ 0; & \text{otherwise} \end{cases} \qquad (1)$$

Instead of binary weights, feature weights can also be used to represent a user session. These feature weights may be based on frequency of occurrence of a URL reference within the user session, the time a user spends on a particular page or the number of bytes downloaded by the user from a page.

## III. DATA CLUSTERING TECHNIQUES

In this section a detailed discussion of each clustering technique and its underlying mathematical model is presented.

### A. *k-Means Clustering Algorithm:*

The *k*-Means clustering or Hard *c*-Means clustering algorithm [16] is one of the most commonly used methods for partitioning the data. Given a set of *m* data points $X = \{x_i | i = 1 \cdots m\}$, where each data point is a *n*-dimensional vector, *k*-means clustering algorithm aims to partition the *m* data points into *k* clusters ($k \leq m$) $C = \{c_1, c_2, ..., c_k\}$ so as to minimize an *objective function* (or a cost function) *J(V, X)* of dissimilarity [17], which is the within-cluster sum of squares. In most cases the dissimilarity measure is chosen as the Euclidean distance. The objective function is an indicator of the distance of the *n* data points from their respective cluster centers. The objective function *J*, based on the Euclidean distance between a data point vector $x_i$ in cluster *j* and the corresponding cluster center $v_j$, is defined in (2).

$$J(X,V) = \sum_{j=1}^{k} J_i(x_i, v_j) = \sum_{j=1}^{k}\left(\sum_{i=1}^{m} u_{ij}.d^2(x_i, v_j)\right), \qquad (2)$$

where, $J_i(x_i, v_j) = \sum_{i=1}^{m} u_{ij}.d^2(x_i, v_j)$,

is the objective function within cluster $c_i$,

$u_{ij} = 1$, if $x_i \in c_j$ and 0 otherwise.

$d^2(x_i, v_j)$ is the disatnce between $x_i$ and $v_j$

Euclidian distance between various data points and cluster centers can be calculated using (3).

$$d^2(x_i, v_j) = \left\| \sum_{k=1}^{n} x_k^i - v_k^j \right\|^2 \quad (3)$$

where, $n$ is the number of dimensions of each data point

$x_k^i$ is the value of $k^{th}$ dimensions of $x_i$

$v_k^j$ is the value of $k^{th}$ dimensions of $v_j$

The k-means clustering first initializes the cluster centers randomly. Then each data point $x_i$ is assigned to some cluster $v_j$ which has the minimum distance with this data point. Once all the data points have been assigned to clusters, cluster centers are updated by taking the weighted average of all data points in that cluster. This recalculation of cluster centers results in better cluster center set. The process is continued until there is no change in cluster centers.

The partitioned clusters are defined by a $m \times k$ binary membership matrix $U$, where the element $u_{ij}$ is 1, if the $i$th data point $x_i$ belongs to the cluster $j$, and 0 otherwise. Once the cluster centers $V = \{v_1, v_2, \ldots, v_k\}$, are fixed, the membership function $u_{ij}$ that minimizes (2) can be derived as follows:

$$u_{ij} = \begin{cases} 1; & \text{if } d^2(x_i, v_j) \le d^2(x_i, v_{j*}) \ j \ne j*, \forall \ j* = 1, \cdots, k \\ 0; & \text{otherwise} \end{cases} \quad (4)$$

The equation (4) specifies that assign each data point $x_i$ to the cluster $c_j$ with the closest cluster center $v_j$. Once the membership matrix $U=[u_{ij}]$ is fixed, the optimal center $v_j$ that minimizes (2) is the mean of all the data point vectors in cluster $j$:

$$v_j = \frac{1}{|c_j|} \sum_{i, x_i \in c_j}^{m} x_i \quad (5)$$

where,

$|c_j|$, is the size of cluster $c_j$ and also $|c_j| = \sum_{i=1}^{m} u_{ij}$

Given an initial set of $k$ means or cluster centers, $V = \{v_1, v_2, \ldots, v_k\}$, the algorithm proceeds by alternating between two steps: i) Assignment step: Assign each data point to the cluster with the closest cluster center. ii) Update step: Update the cluster center as the mean of all the data points in that cluster. The input to the algorithm is a set of $m$ data points $X = \{x_i \mid i = 1 \cdots m\}$, where each data point is a $n$-dimensional vector, it then determines the cluster centers $v_j$ and the membership matrix $U$ iteratively as explained in Fig. 1.

The *k*-means algorithm provides locally optimal solutions with respect to the sum of squared errors represented by the error objective function. Since it is a fast iterative algorithm, it has been applied to a variety of areas [18]-[20].

The attractiveness of the k-means lies in its simplicity and flexibility. However, it suffers from major shortcomings that have been a cause for it not being implemented on large datasets. The most important among these are i) *k*-Means scales poorly with respect to the time it takes for large number of points; ii) The algorithm might converge to a solution that is a local minimum of the objective function. The main disadvantage of this algorithm lies in its sensitivity to initial positions of the cluster centroids [21].

---

**Algorithm:** *k*-Means clustering algorithm for partitioning, where each cluster's center is represented by the mean value of the data points in that cluster.

**Input:** *k*, the number of clusters and Set of *m* data points $X=\{x_1, \ldots, x_m\}$.

**Output:** Set of *k* centroids, $V=\{v_1, \ldots, v_k\}$, corresponding to the clusters $C=\{c_1, \ldots, c_k\}$, and membership matrix $U=[u_{ij}]$.

**Steps:**

4) Initialize the *k* centroids $V=\{v_1, \ldots, v_k\}$, by randomly selecting *k* data points from *X*.

5) **repeat**

   iv) Determine the membership matrix $U$ using (8), by assigning each data point $x_i$ to the closest cluster $c_j$.

   v) Compute the objective function $J(X,V)$ using (6). Stop if it below a certain threshold ε.

   vi) Recompute the centroid of each cluster using (9).

6) **until** Centroids do not change

---

Figure 1. *k*-Means Clustering Algorithm

Since the performance of the k-Means algorithm depends on the initial positions of the cluster centeroids, it is recommended to execute the algorithm multiple times, each with a different set of initial centroids.

*B. K-Medoids Clustering Algorthm:*

*k*-Medoid is a classical partitioning technique of clustering that clusters the data set of *m* data points into *k* clusters. It attempts to minimize the squared error, which is the distance between data points within a cluster and a point designated as the center of that cluster. In contrast to the *k*-means algorithm, *k*-Medoids algorithm selects data points as cluster centers (or medoids). A medoid is a data point of a cluster, whose average dissimilarity to all the other data points in the cluster is minimal i.e. it is a most centrally located data point in the cluster [20],[22].

Given a set of *m* data points $X = \{x_i \mid i = 1 \cdots m\}$, where each data point is a *n*-dimensional vector, *k*-mdoids clustering algorithm aims to partition the *m* data points into *k* clusters ($k \le m$) $C = \{c_1, c_2, \ldots, c_k\}$ so as to minimize an objective function representing the sum of the dissimilarities between each of the data points and its corresponding cluster medoid. Let $M = \{m_1, m_2, \ldots, m_k\}$ be the set of medoids corresponding to *C*. The objective function $J(X, M)$ is defined in (7)

$$J(X,M) = \sum_{j=1}^{k}\left(\sum_{i=1}^{m} u_{ij} \cdot d^2(x_i, m_j)\right), \quad (7)$$

where,

$x_i$ is the $i^{th}$ data point

$m_j$ is the medoid of cluster $c_j$

$u_{ij} = 1$, if $x_i \in c_j$ and 0 otherwise.

$d^2(x_i, m_j)$ is the Euclidean disatnce between $x_i$ and $m_j$

$$d^2(x_i, m_j) = \left\|\sum_{k=1}^{n} x_k^i - m_k^j\right\|^2 \quad (8)$$

where, $n$ is the number of dimensions of each data point

$x_k^i$ is the value of $k^{th}$ dimensions of $x_i$

$m_k^j$ is the value of $k^{th}$ dimensions of $m_j$

The partitioned clusters are defined by a $m \times k$ binary membership matrix $U$, where the element $u_{ij}$ is 1, if the $i$th data point $x_i$ belongs to the cluster $j$, and 0 otherwise. Once the cluster medoids $M = \{m_1, m_2, …, m_k\}$, are fixed, the membership function $u_{ij}$ that minimizes (7) can be derived as follows:

$$u_{ij} = \begin{cases} 1; & \text{if } d^2(x_i, m_j) \leq d^2(x_i, m_{j^*}) \ j \neq j^*, \forall \ j^* = 1, \cdots, k \\ 0; & \text{otherwise} \end{cases} \quad (9)$$

The equation (9) specifies that assign each data point $x_i$ to the cluster medoid $m_j$. Once the membership matrix $U=[u_{ij}]$ is fixed, the new cluster medoids $m_j$ that minimizes (7) can be found using (10)

$$m_j = \arg\min_{x_i \in c_j} \sum_{x_l \in c_j} d(x_i, x_l) \quad (10)$$

The basic strategy of *k*-Medoids clustering algorithms is to discover *k* clusters in *m* objects by first arbitrarily selecting a representative data point (the Medoid) as the center for each cluster. Each remaining data point is clustered with the medoid to which it is the most similar. The algorithm takes the input parameter *k*, the number of clusters to be partitioned among a set of *m* objects.

The most common realisation of *k*-medoid clustering is the Partitioning Around Medoids (PAM) algorithm and is as described in Fig.

It is more robust to noise and outliers as compared to *k*-means because because a medoid is less influenced by outliers or other extreme values than a mean. It minimizes the sum of pair-wise dissimilarities instead of a sum of squared Euclidean distances as in case of k-means. Both methods require the user to specify *k*, the number of clusters.

---

**Algorithm:** *k*-Medoids Clustering

**Input:** Set of *m* data points $X=\{x_1, …, x_m\}$.

**Output:** Set of *k* medoids, $M=\{m_1, …, m_k\}$, corresponding to the clusters $C=\{c_1, …, c_k\}$, and membership matrix $U=[u_{ij}]$ that minimizes the sum of the dissimilarities of all the data points to their nearest medoid.

**Steps:**

1) Initialize the *k* medoids $V=\{v_1, …, v_k\}$, by randomly selecting *k* data points from *X*.

2) **repeat**
   i) Determine the membership matrix *U* using (9), by assigning each data point $x_i$ to the closest cluster $c_j$.
   ii) Compute the objective function $J(X,M)$ using (7). Stop if it below a certain threshold ε.
   iii) Recompute the medoid of each cluster using (10).

3) **until** Medoids do not change

Figure 2. *k*-Medoids Clustering Algorithm

*C. Leader Clustering Algorthm:*

The leader clustering algorithm [23],[24] is based on a predefined dissimilarity threshold. Initially, a random data point from the input data set is selected as leader. Subsequently, distance of every other data point with the selected leader is computed. If the distance of a data point is less than the dissimilarity threshold, that data point falls in the cluster with the initial leader. Otherwise, the data point is identified as a new leader. The computation of leaders is continued till all the data points are considered. It should be noted that the result of the clustering depends on the chosen distance threshold. The number of leaders is inversely proportional to the selected threshold.

Given a set of *m* data points $X = \{x_i \mid i = 1 \cdots m\}$, where each data point is a *n*-dimensional vector. The Euclidean distance between the $i^{th}$ data point $x_i \in X$ and $j^{th}$ leader $l_j \in L$ (where *L* is a set of leaders) is given by:

$$d^2(x_i, l_j) = \left\|\sum_{k=1}^{n} x_k^i - l_k^j\right\|^2 \quad (11)$$

where, $n$ is the number of dimensions of each data point

$x_k^i$ is the value of $k^{th}$ dimensions of $x_i$

$l_k^j$ is the value of $k^{th}$ dimensions of $x_j$

Fig. 3 below describes the leader clustering algorithm

```
Algorithm: Leader Clustering
Input:   i) Set of m data points X={x_1, ..., x_m},
         ii) α, the dissimilarity threshold.
Output:  Set of clusters C = {c_1, ..., c_k},
Steps:
1)  C = φ, L = φ, j = 1    // Initilize the cluster and leader sets
2)  l_j = x_1              // Initialize x_1 as the first leader
3)  L = L ∪ l_j
4)  c_j = c_j ∪ x_1
5)  C = C ∪ c_j
6)  for each  x_i ∈ X  where i = 2, ... m
7)  begin
8)      j = arg min d(x_i, l_j)
             j, lj∈L
9)      if  d^2(x_i, l_j) < α then
10)         c_j = c_j ∪ x_i
11)     else
12)         j = j + 1
13)         l_j = x_i
14)         L = L ∪ l_j
15)         c_j = c_j ∪ x_i
16)         C = C ∪ c_j
17)     endif
18) end
```

### D. DBSCAN Clustering Algorthm:

DBSCAN (Density-Based Spatial Clustering of Applications with Noise) [25] is a density-based data clustering algorithm because it finds a number of clusters starting from the estimated density distribution of corresponding nodes.

Figure 3. Leader Clustering Algorithm

Given a set of $m$ data points $X = \{x_i \mid i = 1 \cdots m\}$, where each data point is a $n$-dimensional vector. The Euclidean distance between the two data points $x_p \in X$ and $x_q \in X$ is given by

$$d^2(x_p, x_q) = \left\| \sum_{k=1}^{n} x_k^p - x_k^q \right\|^2 \qquad (12)$$

where, $n$ is the number of dimensions of each data point
$x_k^p$ is the value of $k^{th}$ dimensions of $x_p$
$m_k^q$ is the value of $k^{th}$ dimensions of $x_q$

In this algorithm concept of a cluster is based on the notion of "ε-neighborhood" and "density reachability". Let the ε-neighborhood of a data point $x_p$, denoted as $N_\varepsilon(x_p)$ is defined as below:

$$N_\varepsilon(x_p) = \left\{ x_q \in X \mid d^2(x_p, x_q) \leq \varepsilon \right\} \qquad (13)$$

where, $\varepsilon$ is the neighborhood distance

Let $\eta$ be the minimum number of points required to form a cluster. A point $x_q$ is directly density-reachable from a point $x_p$, if $x_q$ is part of ε-neighborhood of $x_p$ and if the number of points in the ε-neighborhood of $x_p$ are greater than or equal to $\eta$ as specified in (13)

$$x_q \in N_\varepsilon(x_p) \qquad (14)$$
$$|N_\varepsilon(x_p)| \geq \eta$$

where $\eta$ is the minimum number of points required for a cluster

$x_q$ is called density-reachable from $x_p$ if there is a sequence $x_1, \ldots, x_n$ of points with $x_1 = x_p$ and $x_n = x_q$ where each $x_{i+1}$ is directly density-reachable from $x_i$. Two points $x_p$ and $x_q$ are said to be density-connected if there is a point $x_o$ such that $x_o$ and $x_p$ as well as $x_o$ and $x_q$ are density-reachable.

A cluster of data points satisfies two properties: i) All the data points within the cluster are mutually density-connected. ii) If a data point is density-connected to any data point of the cluster, it is part of the cluster as well.

Input to DBSCAN algorithm are i) ε (epsilon) and ii) $\eta$, the minimum number of points required to form a cluster. The algorithm starts by randomly selecting a starting data point that has not been visited. If the ε-neighborhood of this data point contains sufficiently many points, a cluster is started. Otherwise, the data point is labeled as noise. Later this point might be found in a sufficiently sized ε-neighborhood of a different data point and hence could become part of a cluster. If a data point is found to be part of a cluster, all the data points in its ε-neighborhood are also part of that cluster and hence added to the cluster. This process continues until the cluster is completely found. Then, a new unvisited point is selected and processed, leading to the discovery of a next cluster or noise. Fig. 4 describes the DBSCAN algorithm.

Although DBSCAN can cluster objects given input parameters such as ε and $\eta$, but it is the responsibility of the user to select these parameter values. Such parameter settings are usually empirically set and difficult to determine, especially for high-dimensional data sets.

```
Algorithm: DBSCAN
Input:    i) Set of m data points X={x₁, …, xₘ},
         ii) ε (epsilon), the neighborhood distance and
        iii) η , the minimum number of data points
             required to form a cluster.
Output:  Set of clusters C = {c₁, …, cₖ},
Steps:
1)  C = Ø.; i = 0;
2)  for each  xₚ ∈ X and xₚ.visited = false
3)  begin
4)      xₚ.visited = true
5)      Nₚ = N_ε(xₚ) using (13)
6)      if |N_ε(xₚ)| < η then
7)          xₚ.noice = true
8)      else
9)          i = i + 1
10)         C = C ∪ cᵢ
11)         cᵢ = cᵢ ∪ xₚ
12)         for each  x_q ∈ N
13)         begin
14)             if x_q.visited = false then
15)                 x_q.visited = true
16)                 N_q = N_ε(x_q)
17)                 if |N_ε(x_q)| < η then
18)                     Nₚ = Nₚ ∪ N_q
19)                     if x_q ∉ c_j ∀j = 1 ≤ j ≤ i then
20)                         cᵢ = cᵢ ∪ x_q
21)                     endif
22)                 endif
23)             endif
24)         end
25)     endif
26) end
```

Figure 4. DBSCAN Algorithm

## IV. EXPERIMENTAL RESULTS

In order to discover the clusters that exist in user accesses sessions of a web site, we carried out a number of experiments using various clustering techniques. The Web access logs were taken from the P.A. College of Engineering, Mangalore web site, at URL http://www.pace.edu.in. The site hosts a variety of information, including departments, faculty members, research areas, and course information. The Web access logs covered a period of one month, from February 1, 2011 to February 8, 2011. There were 12744 logged requests in total.

### A. Preprocessing the Web Log Data:

After performing the cleaning operation the output file contained 12744 entries. Total numbers of unique users identified are 16 and the number of user sessions discovered are 206. Table II depicts the results of cleaning and user identification and user session identification steps of preprocessing. Further details of our preprocessing approaches can be found from our previous work [26].

TABLE II

RESULTS OF CLEANING AND USER IDENTIFICATION

| Items | Count |
|---|---|
| Initial No of Log Entries | 12744 |
| Log Entries after Cleaning | 11995 |
| No. of site ULRs accessed | 116 |
| No of Users Identified | 16 |
| No. of User Sessions Identified | 206 |

Figure 4 shows the result of user session identification. It depicts the percentage of user sessions accessing the specified number of URLs.

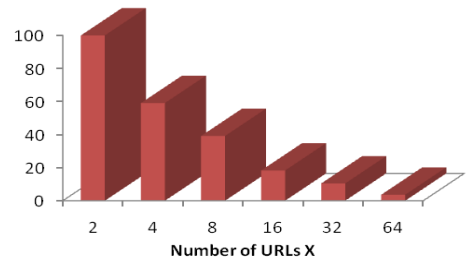

Figure 5. Percentage of Sessions accessing X No. of URLs

### B. Clustering of User Navigational Sessions:

Once the user sessions are discovered, user session data is presented to the four different clustering algorithms in order to discover session clusters that represent similar URL access patterns. These algorithms are i) *k*-Means ii) *k*-Medoids iii) Leader and iv) DBSCAN. Since the above clustering algorithms result in different clusters it is important to perform an evaluation of the results to assess their quality. We evaluated our results based on DB index and C Index which are two quality measures to evaluate the quality of the discovered clusters. These validity measures a described below:

*Davies-Bouldin Validity Index*: This index attempts to minimize the average distance between each cluster and the one most similar to it. It is defined as:

$$DB = \frac{1}{k} \sum_{i=1}^{k} \max_{1 \le j \le k, j \ne i} \left( \frac{diam(c_i) + diam(c_j)}{dis(c_i, c_j)} \right) \quad (16)$$

An optimal value of the k is the one that minimizes this index.

*C Index*: It is defined as [28]:

$$C = \frac{S - S_{min}}{S_{max} - S_{min}}, \quad (17)$$

Here *S* is the sum of distances over all pairs of objects form the same cluster. Let *m* be the number of those pairs and $S_{min}$ is the sum of the *m* smallest distances if all pairs of objects are considered. Similarly $S_{max}$ is the sum of the *m* largest distances out of all pairs. The interval of the C-index values is [0, 1] and this value should be minimized.

The results of application of various clustering algorithms are presented in the following subsections.

*1) k-Means Algorithm:*

We conducted multiple runs of k-Means algorithm by selecting the input parameter *k* (number of clusters) ranging from k = 2, …, 67. (The value 67 for the number of clusters is one third of total number of the discovered user sessions). For each of these runs we computed the value of the clustering error function (*J*) using (2) which represents the sum of the squared error. We also computed the execution timings, Dunn's index, DB index and C index for all of the above runs. Table III describes the results after the application of *k*-Means clustering algorithm.

TABLE III

K-MEANS CLUSTERING RESULTS

| Clusters | SSE (*J*) | DB Index | C Index | Execution Time(ms) |
|---|---|---|---|---|
| 10 | 583.54 | 1.3395 | 0.1229 | 49 |
| 20 | 443.06 | 1.3456 | 0.1060 | 110 |
| 30 | 357.24 | 1.2228 | 0.0769 | 142 |
| 40 | 284.08 | 1.1045 | 0.0610 | 164 |
| 50 | 279.29 | 1.1345 | 0.0651 | 278 |
| 60 | 260.64 | 0.8846 | 0.0783 | 188 |

One of the problems associated with the k-Means algorithm is that it may produce empty clusters depending on the initial centroids chosen. Graph in Fig. 6 Describes the number of empty clusters generated for different values of *k*. M.K. Pakhira [29] has proposed a modified k-means algorithm to avoid the empty clusters. K-medoids algoritm also rectifies this problem.

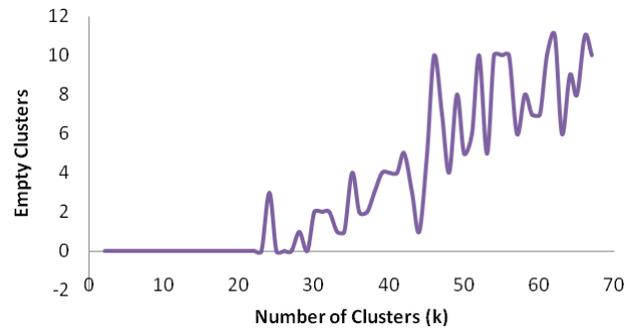

Figure 6. No. of Empty Clusters Vs. No. of Initial Clusters *k*

*2) k-Medoids Algorithm:*

We conducted the multiple runs of *k*-Medoids algorithm by selecting the input parameter *k* (number of clusters) ranging from k = 2, …, 67. (The value 67 for the number of clusters is one third of total number of the discovered user sessions). For each of these runs we computed the value of the clustering error function (*J*) using (7), which represents the sum of the squared error. We also computed the execution timings, Dunn's index and DB index and C index for all of the above runs. Table IV describes the results after the application of *k*-Means clustering algorithm.

TABLE IV

K-MEDOIDS CLUSTERING RESULTS

| Clusters | Error (*J*) | DB Index | C Index | Execution Time(ms) |
|---|---|---|---|---|
| 10 | 613.73 | 1.4426 | 0.1622 | 7 |
| 20 | 512.81 | 1.4689 | 0.1543 | 7 |
| 30 | 352.88 | 1.2018 | 0.05 | 5 |
| 40 | 315.63 | 0.9413 | 0.23572 | 6 |
| 50 | 257.83 | 2.35 | 0.03 | 7 |
| 60 | 254.13 | 2.85 | 0.06 | 9 |

We compared the k-Means and k-Medoids algorithms based on clustering error (*J* as defined in equations (2) and (7)), cluster validity using C index and the execution time.

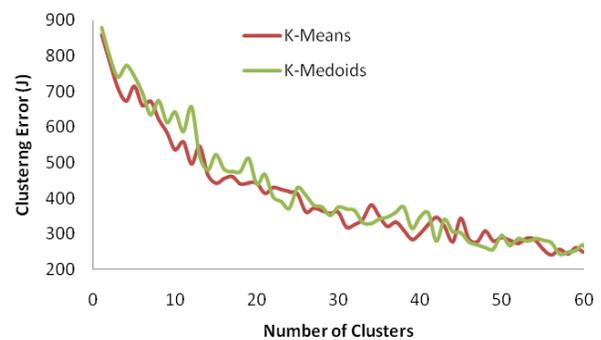

Figure 7. Clustering Error (*J*) Vs. No. of Clusters *k*

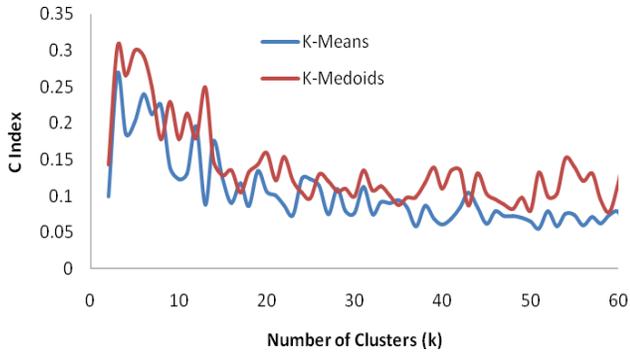

Figure 8.  C Index Vs. No. of Clusters *k*

Our results (Fig. 7) show that the k-Means algorithm minimizes the clustering error (*J*) slightly better than the k-Medoids algorithm. C index values in graph plot of Fig. 8 indicates that the clusters of *k*-Means algorithm have better validity index than that of *k*-Medoids algorithm. On the other the execution timings of *k*-Medoids algorithms are faster than the that of *k*-Means algorithm as show in Fig.9.

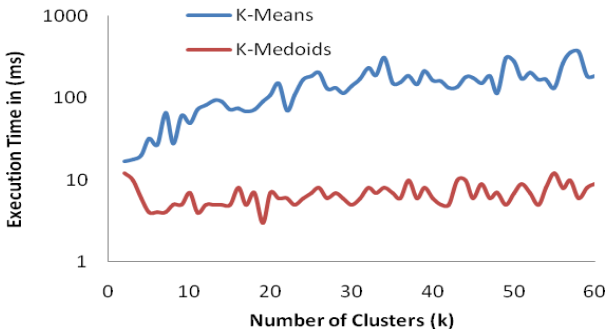

Figure 9.  Execution Time in milliseconds Vs. No. of Clusters *k*

### 3) Leader Algorithm:

We conducted the multiple runs of Leader algorithm by selecting the input parameter ε (Dissimilarity Threshold) ranging from ε = 0.5, …, 3.5 in steps of 0.5. For each of these runs we computed the value of the clustering error. We also computed the execution timings, DB index and C index for all of the above runs. Table V describes the results after the application of Leader clustering algorithm.

TABLE V

LEADER CLUSTERING RESULTS

| Epsilon (ε) | Error (*J*) | DB Index | C Index | Execution Time(ms) | No. of Clusters |
|---|---|---|---|---|---|
| 1 | 26.19 | 0.3623 | 0.0021 | 3 | 115 |
| 1.5 | 76.81 | 0.5061 | 0.0348 | 2 | 86 |
| 2 | 216.62 | 0.5578 | 0.0588 | 2 | 56 |
| 2.5 | 398.81 | 0.7200 | 0.0801 | 1 | 33 |
| 3 | 467.07 | 0.9084 | 0.1878 | 2 | 26 |
| 3.5 | 624.87 | 0.8801 | 0.2407 | 1 | 14 |

Fig. 10 shows the results of Leader clustering. From the graph it is very clear that the number of discovered clusters is inversely proportional to the dissimilarity threshold ε.

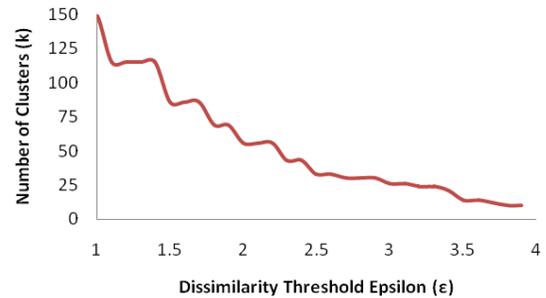

Figure 10.  Number of clusters formed Vs. Dissimilarity Threshold ε

### 4) DBSCAN Algorithm:

We conducted the multiple runs of DBSCAN algorithm by selecting the input parameter ε (neighborhood disatnace) ranging from ε = 0.5, …, 3.5 in steps of 0.5. The other parameter η which indicates the minimum no. of points in a cluster is set in a range from η = 2, …, 10. For each of these runs we computed the value of the clustering error. We also computed the execution timings, DB index and C index for all of the above runs. Table VI describes the results after the application of DBSCAN algorithm for the value of η = 2.

TABLE VI

DBSCAN RESULTS

| Epsilon (ε) | Error (*J*) | DB Index | C Index | Execution Time(ms) | No. of Clusters |
|---|---|---|---|---|---|
| 1 | 766.9 | 1.2594 | 0.6606 | 13 | 21 |
| 1.5 | 805.881 | 1.3665 | 0.1984 | 20 | 7 |
| 2 | 871.2758 | 0.8415 | 0.0766 | 24 | 2 |
| 2.5 | 881.5672 | 0.8348 | 0.0500 | 13 | 2 |
| 3 | 866.1479 | 1.0874 | 0.0442 | 16 | 3 |
| 3.5 | 867.23 | 0.9092 | 0.0463 | 17 | 3 |

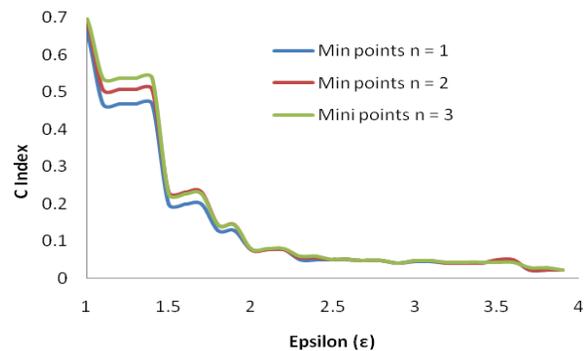

Figure 11.  C Index Vs. Neighbourhood Distance ε

The graph plot in Fig. 11 displays the C index as a function of the neighbourhood distance ε, for different values of η (the minimum number of points in a cluster). The graph shows that the C index value improves as we increase the neighbourhood distance ε. It also improves if we decrease the value of η.

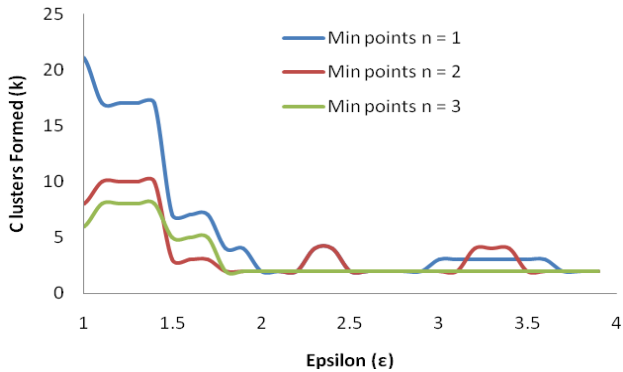

Figure 12. Number of clusters formed Vs. Neighbourhood Distance ε

The graph plot in Fig. 12 displays the number of clusters formed as a function of the neighbourhood distance ε, for different values of η (the minimum number of points in a cluster). The graph shows that the number of clusters formed decreases as we increase the neighbourhood distance ε. It also decreases if we increase the value of η.

The next two graphs compare the results of the Leader and DBSCAN techniques.

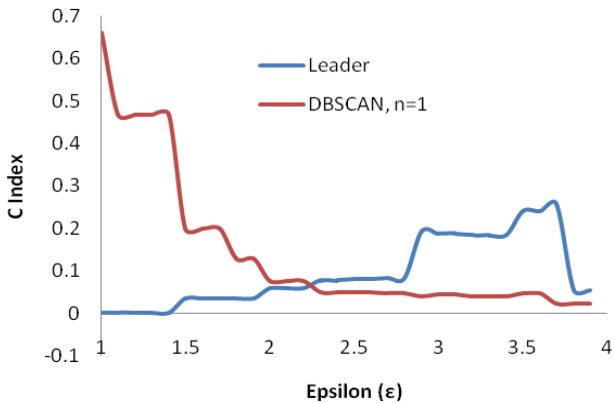

Figure 13. C Index Vs. Epsilon (ε)

The graph plot in Fig. 13 displays the C validity index value as a function of Epsilon (ε). Here ε is the dissimilarity threshold in case of Leader clustering and neighbourhood distance in case of DBSCAN. Our results show that in case of Leader clustering, validity index improves for lower values of dissimilarity distance ε. In case of DBSCAN, the validity index improves as increase the value of neighbourhood distance ε. Note that we have set the value of η to 1.

The graph plot in Fig.14 displays the Execution Time as a function of Epsilon (ε). It is clear from the graph that the Leader algorithm performs much faster than the DBSCAN if we keep the Leader dissimilarity threshold and DBSCAN neighbourhood distance same. Note that we have set the value of η to 1.

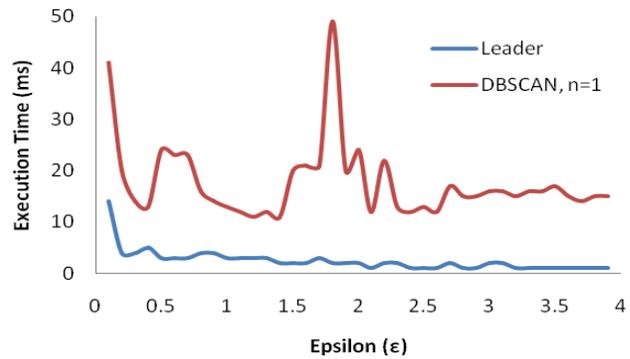

Figure 14. Execution Time in milliseconds Vs. Epsilon (ε)

I. CONCLUSION AND FUTURE WORK

In this paper we have presented our framework for web usage data clustering for users' navigational sessions using k-Means, k-Medoids, Leader and DBSCAN clustering algorithms. We provided a detailed overview of these techniques. We also described the mathematical model and algorithm details related to the implementation of these clustering algorithms in order to discover the user sessions clusters. From the results presented in the previous section, we conclude the following points.

- K-means clustering produces fairly higher accuracy and lower clustering error as compared with k-medoids clustering algorithm.

- K-means algorithm may result in the formation of empty cluster while it is not the case with k-medoids algorithm.

- Our result shows that k-medoids algorithm gives reasonably better time performance than that of the k-means algorithm. The reason behind this is we are using a large data set. The k-Medoids algorithm requires to compute the distance between every pair of data objects only once and uses this distance at every stage of iteration. On the other for an optimal solution k-Means algorithm performs multiple runs and computes the distance between every data object and it's corresponding cluster center.

- Although Leader clustering algorithm does not require estimating the value of *k* at the beginning, it does require estimating the dissimilarity threshold ε.

- Number of clusters formed in Leader clustering is inversely proportional to the value of dissimilarity threshold ε.

- Leader clustering validity index (C index) improves as we increase the value of the dissimilarity threshold ε.

- DBSCAN algorithm can identify a data point as a noise or outlier.

- DBSCAN validity index (C index) improves as we decrease the value of the neighborhood distance ε.

- If we choose the same value for dissimilarity threshold in Leader clustering and neighbor distance in DBSCAN (while keeping η constant), the time performance of Leader clustering much faster than that of DBSCAN.

Another direction of future work is related with the use of fuzzy *c*-Mean clustering technique to discover the user session clusters. The reason behind this is, although the several clustering algorithms described are suitable in handling the crisp data which have clear cut boundaries, but in reality web usage data is semi-structured and contains the outliers and incomplete navigational data, due to a wide variety of reasons inherent to web browsing and logging. Therefore, Web Usage Mining requires modelling of multiple overlapping sets in the presence of significant noise and outliers. Soft Computing based techniques such as Fuzzy Clustering can be very useful for mining such semi structured, noisy and incomplete data.